\documentclass[useAMS,usenatbib,twocolumn,apjl,iop]{emulateapj}
\usepackage{graphicx}
\usepackage{amssymb}
\usepackage{amsmath}
\usepackage{fleqn}

\long\def\symbolfootnote[#1]#2{\begingroup%
\def\thefootnote{\fnsymbol{footnote}}\footnote[#1]{#2}\endgroup} 
\def\apjl{ApJ}
\def\apjs{ApJS}
\def\aap{A\&A}

\def\ges{{Gaia--ESO }}
\def\feh{{\mathrm{[M/H]}}}
\def\kms{{\mathrm{\,kms^{-1}}}}

\def\deg{{\mathrm{deg}}}
\def\degree{\ensuremath{^\circ}}
\def\change#1{{#1}}

\begin{document}
\title{The Gaia-ESO survey: Metal--rich bananas in the bulge}
\author{Angus A. Williams$^1$} \author{N.W.  Evans$^1$}
\author{Matthew Molloy$^2$} \author{Georges Kordopatis$^3$}
\author{M.C. Smith$^4$} \author{J. Shen$^4$} \author{G. Gilmore$^1$}
\author{S. Randich$^5$} \author{T. Bensby$^6$}
\author{P. Francois$^7$} \author{S.E Koposov$^{1}$}
\author{A. Recio-Blanco$^8$} \author{A. Bayo$^9$}
\author{G. Carraro$^{10}$}, \author{A. Casey$^1$}
\author{T. Costado$^{11}$} \author{E. Franciosini$^5$}
\author{A. Hourihane$^1$} \author{P. de Laverny$^8$}
\author{J. Lewis$^1$} \author{K. Lind$^{12}$} \author{L. Magrini$^5$}
\author{L. Monaco$^{13}$} \author{L. Morbidelli$^5$} \author{G.G
  Sacco$^5$} \author{C. Worley$^1$} \author{S. Zaggia$^{14}$}
\author{\v{S}. Mikolaitis$^{15}$}

\email{aamw3@ast.cam.ac.uk,nwe@ast.cam.ac.uk} \affil{$^1$ Institute of
  Astronomy, Madingley Road, Cambridge CB3 0HA, UK, $^2$ Kavli
  Institute for Astronomy \& Astrophysics, Peking University, Beijing
  100871, China, $^3$ Leibniz-Institut f{\"u}r Astrophysik Potsdam, An
  der Sternwarte 16, 14482 Potsdam, Germany, $^4$ Key Laboratory for
  Research in Galaxies and Cosmology, Shanghai Astronomical
  Observatory, Chinese Academy of Sciences, 80 Nandan Rd, Shanghai
  200030, $^5$ INAF Osservatorio Astrofisico di Arcetri, Largo
  E. Fermi 5, 50125 Florence, Italy, $^6$ Lund Observatory, Dept of
  Astronomy and Theoretical Physics, Box 43, SE-221 00 Lund, Sweden,
  $^7$ GEPI, Observatoire de Paris, CNRS, Universit\'e Paris Diderot,
  5 Place Jules Janssen, 92190 Meudon, France $^8$ Laboratoire
  Lagrange, Universit\'e C\^ote d'Azur, Observatoire de la C\^ote
  d'Azur, CNRS, Bvd de l'Observatoire, CS 34229, 06340 Nice, France,
  $^{9}$ Instituto de F\'isica y Astronom\'ia, Universidad de
  Valpara\'iso, Chile, $^{10}$ European Southern Observatory,Alonso de
  Cordova 3107 Vitacura, Santiago de Chile, Chile, $^{11}$ Instituto
  de Astrof\'isica de Andaluc\'ia-CSIC, Apdo 3004, 18080 Granada,
  Spain, $^{12}$ Max-Planck Institut f{\"u}r Astronomie,
  K{\"o}nigstuhl 17, 69117 Heidelberg, Germany, $^{13}$ Departamento
  de Ciencas Fisicas, Universidad Andres Bello, Republica 220,
  Santiago, Chile, $^{14}$ INAF - Padova Observatory, Vicolo
  dell'Osservatorio 5, 35122 Padova, Italy, $^{15}$ Institute of
  Theoretical Physics and Astronomy, Vilnius University,
  Saul\.{e}tekio al. 3, LT-10222, Vilnius, Lithuania}

\begin{abstract}
We analyse the kinematics of $\sim 2000$ giant stars in the direction
of the Galactic bulge, extracted from the \ges survey in the region
$-10\degree \lesssim \ell \lesssim 10\degree$ and $-11\degree \lesssim
b \lesssim -3\degree$. We find distinct kinematic trends in the metal
rich ($\feh>0$) and metal poor ($\feh<0$) stars in the data. The
velocity dispersion of the metal--rich stars drops steeply with
latitude, compared to a flat profile in the metal--poor stars, as has
been seen previously. We argue that the metal--rich stars in this
region are mostly on orbits that support the boxy--peanut shape of the
bulge, which naturally explains the drop in their velocity dispersion
profile with latitude. The metal rich stars also exhibit peaky
features in their line--of--sight velocity histograms, particularly
along the minor axis of the bulge. We propose that these features are
due to stars on resonant orbits supporting the boxy--peanut
bulge. This conjecture is strengthened through the comparison of the
minor axis data with the velocity histograms of resonant orbits
generated in simulations of buckled bars.  The `banana' or 2:1:2
orbits provide strongly bimodal histograms with narrow velocity peaks
that resemble the \ges metal-rich data.
\end{abstract}

\section{Introduction}

\begin{figure}
\includegraphics[width=.9\columnwidth]{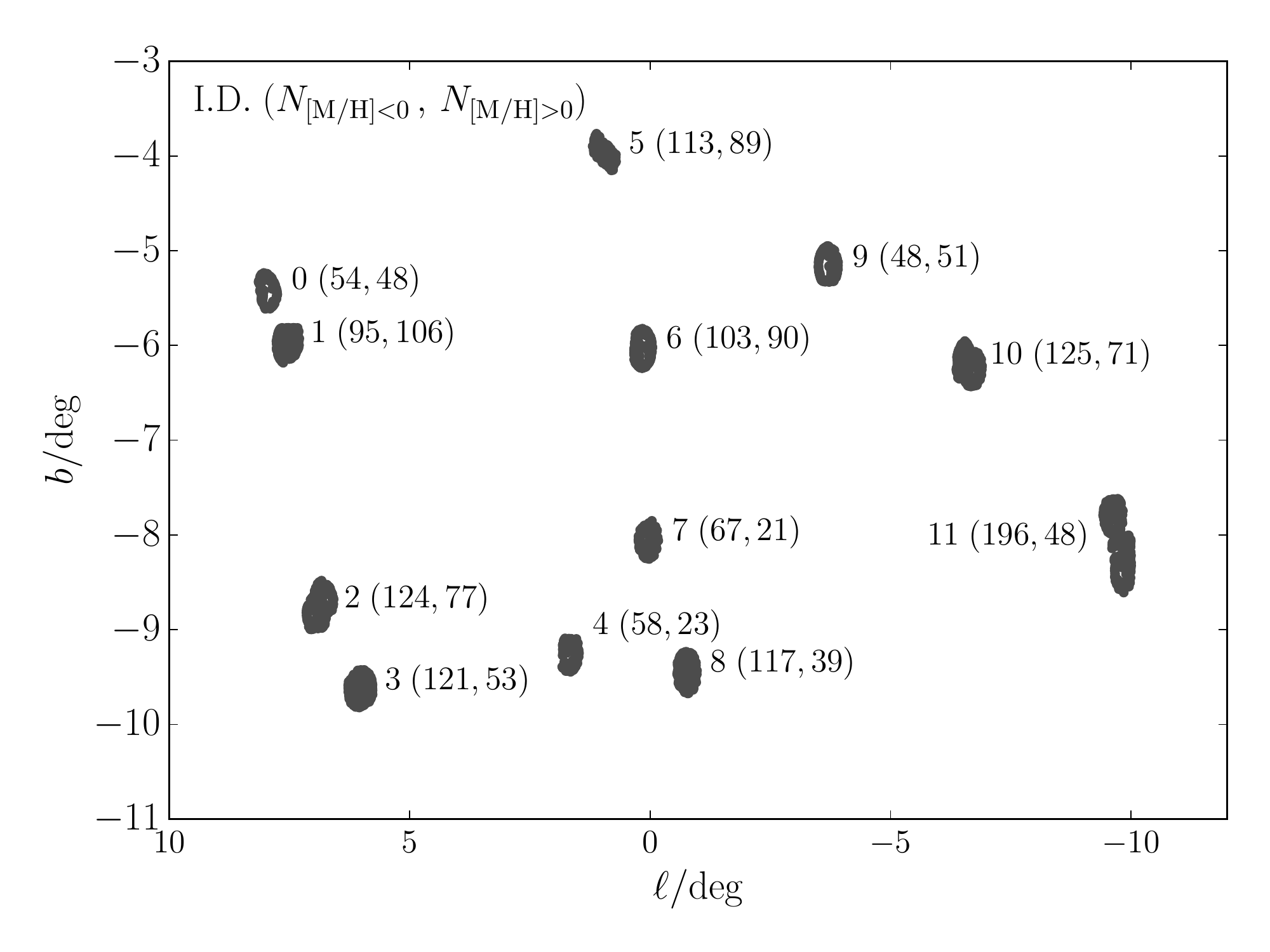}
\caption{The distribution of our sample in Galactic longitude and
  latitude. Each field is labelled by its I.D. number, followed in 
  brackets by the number of stars with $\feh<0$ and the number 
  with $\feh>0$.}
\label{fig:lbdist}
\end{figure}

The metallicity and velocity distributions change across the Galactic
Bulge, as the chemical history and orbital content varies from field
to field~\citep[e.g.,][]{Ba16}.  One of the aims of the large-scale
spectroscopic surveys (such as ARGOS, APOGEE and Gaia-ESO) is to map
out the correlations between metallicity and
kinematics~\citep[e.g.,][]{Ne13,Ro14,Ne15}. This offers the promise of
uncovering the history of the Bulge, as well as its present-day
structure.

Bars are largely built from orbits trapped around the main prograde
periodic orbits aligned with the long axis. These nearly planar orbits
can become vertically unstable and sire families librating around the
vertical resonances~\citep[e.g.,][]{Pf91}. They are called the
`banana' or `pretzel' orbits because of their characteristic
morphology~\citep{Mi89,Po15}. It has long been suspected that they
support the boxy or peanut-shaped bulges seen in external galaxies.
The discovery of the bimodal distribution in red clump magnitudes on
minor axis fields~\citep{McW10,Na10} seemed to provide evidence for
the existence of resonant orbits in the Galactic Bulge
itself~\citep[e.g.,][]{Po15}.

In this {\it Letter}, we examine the bulge fields in the fourth data
release of the Gaia-ESO survey in Section 2. We identify narrow
features in the metallicity and velocity distributions as the
signature of resonant orbits (Section 3) and provide a comparison with
the expected properties of resonant orbits from simulations (Section
4). This is suggestive of the presence of substantial numbers
of banana orbits in the bar, predominantly populated by the metal-rich
stars.

\section{Data}

\begin{figure}
\includegraphics[width=.9\columnwidth]{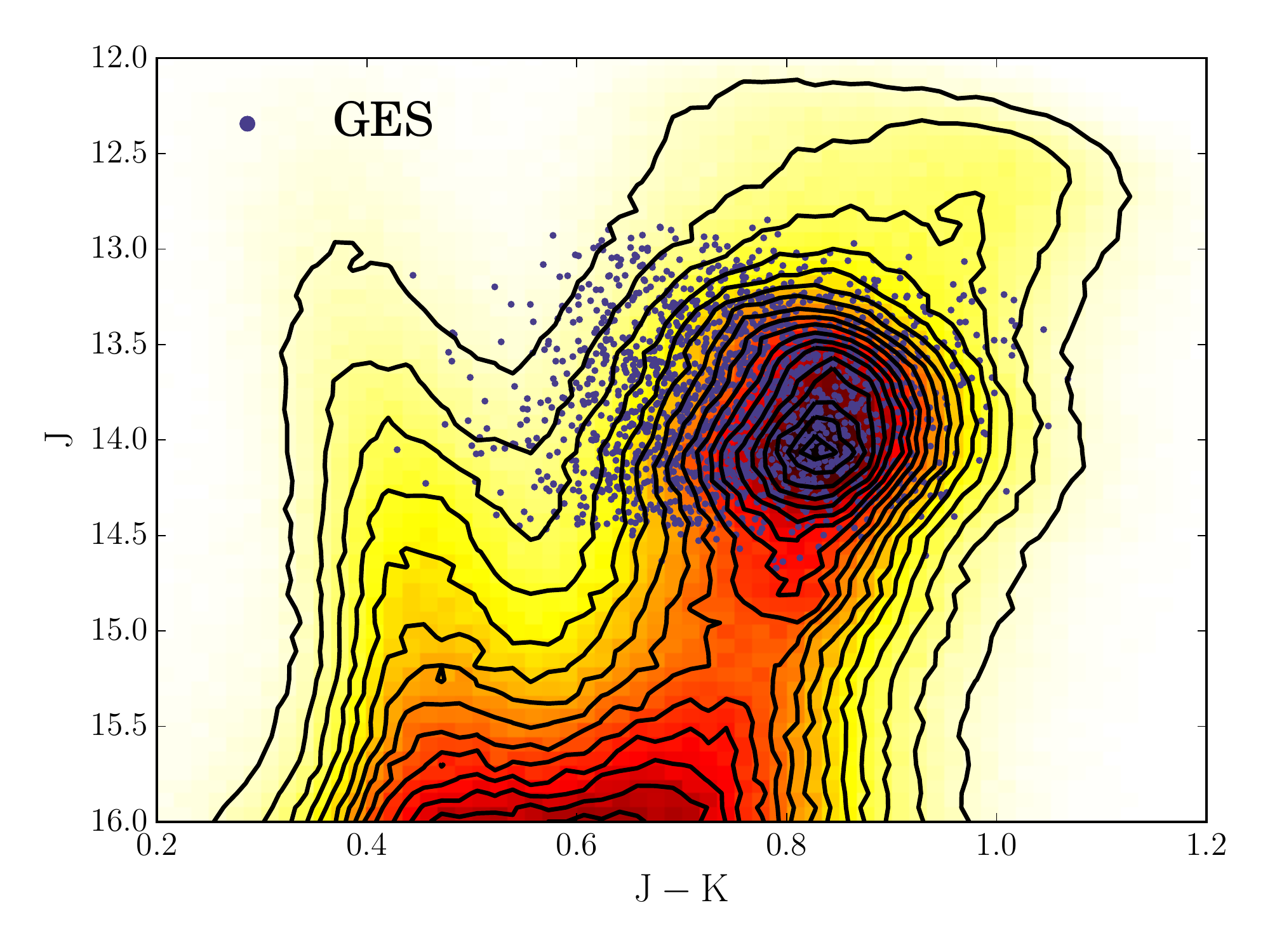}
\caption{Selection of the \ges bulge targets in colour and magnitude.
  The contours are of the combined VVV photometry from all of the
  fields, and the purple points represent our sample.}
\label{fig:CMD}
\end{figure}

\begin{figure}
\includegraphics[width=.9\columnwidth]{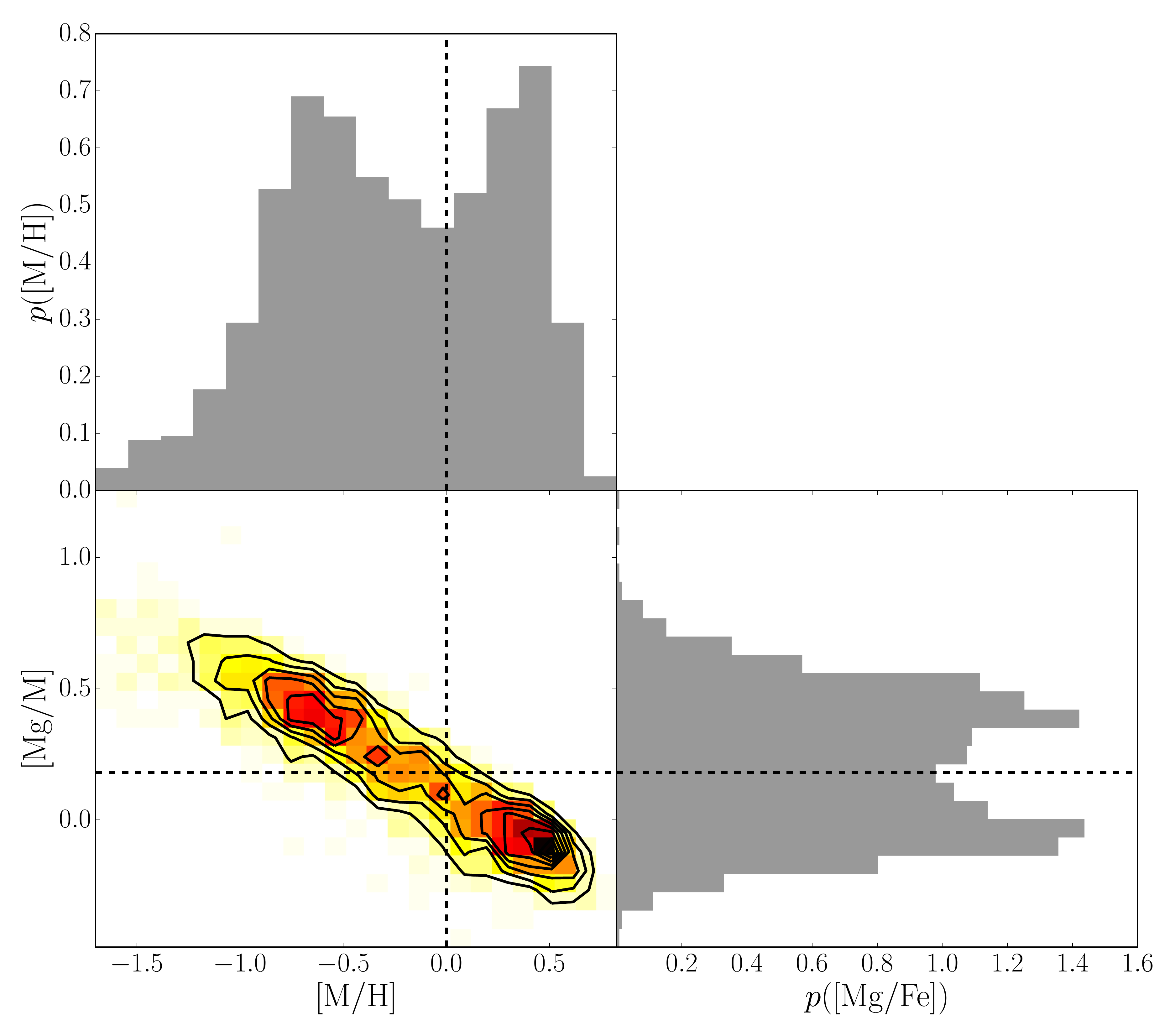}
\caption{$\mathrm{[Mg/M]}$--$\feh$ distribution of our sample. 
  One can clearly see two overdensities in
  this plane, which are reflected in the two marginalised
  distributions. We split the sample around $\feh=0$, which
  effectively also splits in $\mathrm{[Mg/M]}$. The low
  metallicity population has higher $\alpha$--element enrichment,
  whereas the high metallicity population resembles the disc.}
\label{fig:abunplane}
\end{figure}

\ges is a public spectroscopic survey, providing high--quality spectra
for $\sim 10^5$ stars in the Milky Way \citep{Gi12}. Spectra are taken
using the VLT--FLAMES instrument. Here, we analyse a subset of 1938 of
these stars located in the direction of the Galactic bulge, taken from
the 4th internal data release (iDR4). We make a metallicity cut, $-1.7
\leq \mathrm{[M/H]} \leq 1.$, and a surface gravity cut, $\log g <
3.5$, to ensure that we have a clean sample of giants. Our specific
flag cuts are: $\mathrm{GES\textunderscore FLD==``Bulge"}$ and
$\mathrm{GES\textunderscore TYPE==``GE\textunderscore
  MW\textunderscore BL"}$. Typical uncertainty in line--of--sight
velocity (metallicity) is $\sim 0.4\kms$ ($\sim 0.1\,\mathrm{dex}$).

Fig. \ref{fig:lbdist} depicts the distribution of the \ges bulge sample 
(including our cuts) in longitude and latitude, and Fig. \ref{fig:CMD} depicts 
the selection in colour and magnitude with the combined VVV
photometry for all the fields underlaid. The
sample from iDR4 covers a larger area on the sky than previous data
releases \citep[see e.g.][]{Ro14}, giving us more scope to constrain
trends with on--sky position. Each of our fields contains between 80
and 250 stars.  The \ges bulge selection consists of a colour cut in
each field of $\mathrm{(J-K)_0} > 0.38$, which is then reddened
appropriately given the extinction in that direction, and a magnitude
cut $12.9 < \mathrm{J}_0 < 14.1$, which is also modified in some
fields to allow both red clump populations to be represented in the
sample (for more details see \citealt{Ro14}).

\section{Kinematics as a function of metallicity}

\begin{figure}
\includegraphics[width=.9\columnwidth]{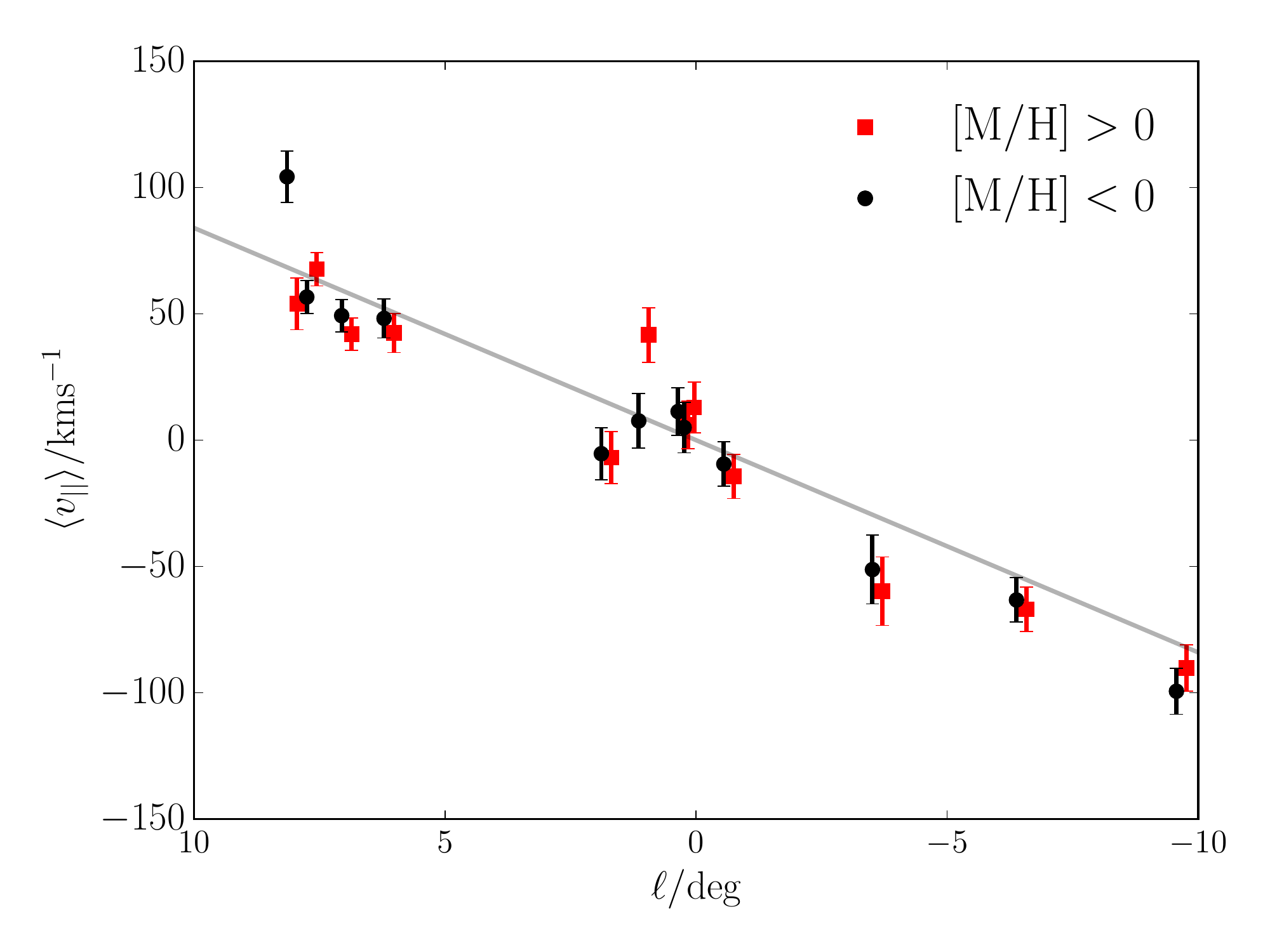}
\caption{Mean line--of-sight velocities a function of longitude. At each longitude, 
  the red squares (metal--rich) and black circles (metal--poor) are slightly offset 
  from one--another for clarity. Both the metal--rich and metal--poor samples exhibit 
  the usual rotational signature seen in bulge stars. The line is a linear fit to the data
  with gradient $8.4\kms\deg^{-1}$. The metal--rich and metal--poor
  rotation curves are both consistent with this value at a $1\sigma$
  level.}
\label{fig:rotation}
\end{figure}

\begin{figure}
\includegraphics[width=.9\columnwidth]{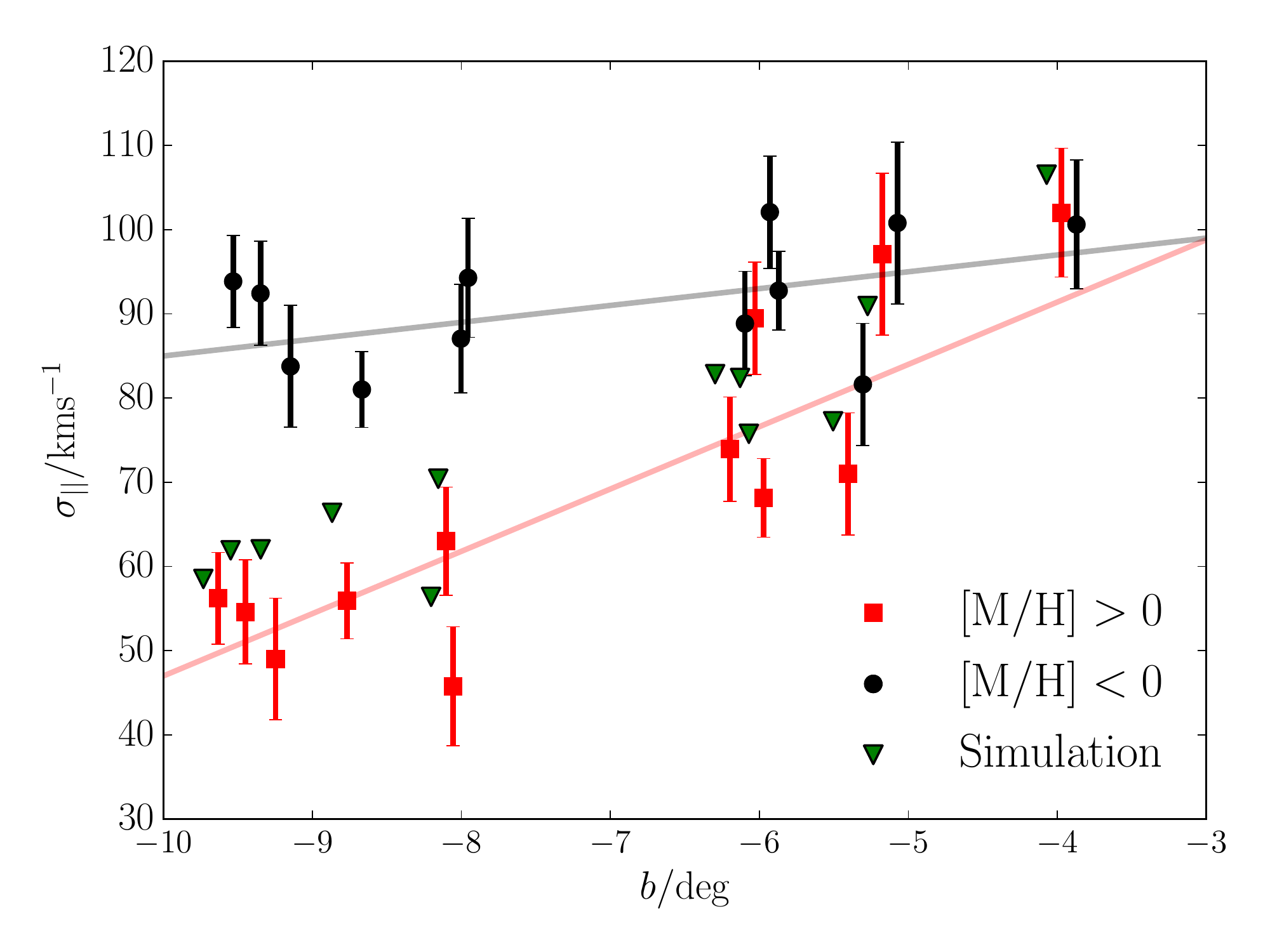}
\caption{Line--of--sight velocity dispersion as a function of
  latitude. At each latitude, the red squares (metal--rich), black circles 
  (metal--poor) and green triangles (simulation) are slightly offset from 
  one--another for clarity. The metal--rich stars in our sample become significantly
  colder as $|b|$ increases, whereas the metal--poor stars are
  consistent with being one temperature. The two lines are simple
  linear fits to the data. The metal--poor stars have a velocity
  dispersion gradient of $2.0\pm 1.0 \kms\deg^{-1}$, whereas the
  metal--rich stars have a gradient of $7.4\pm 0.7 \kms\deg^{-1}$. The
  green triangles show the velocity dispersion extracted from the
  simulation.
}
\label{fig:disps_vs_lat}
\end{figure}

\begin{figure*}
\includegraphics[width=1.9\columnwidth]{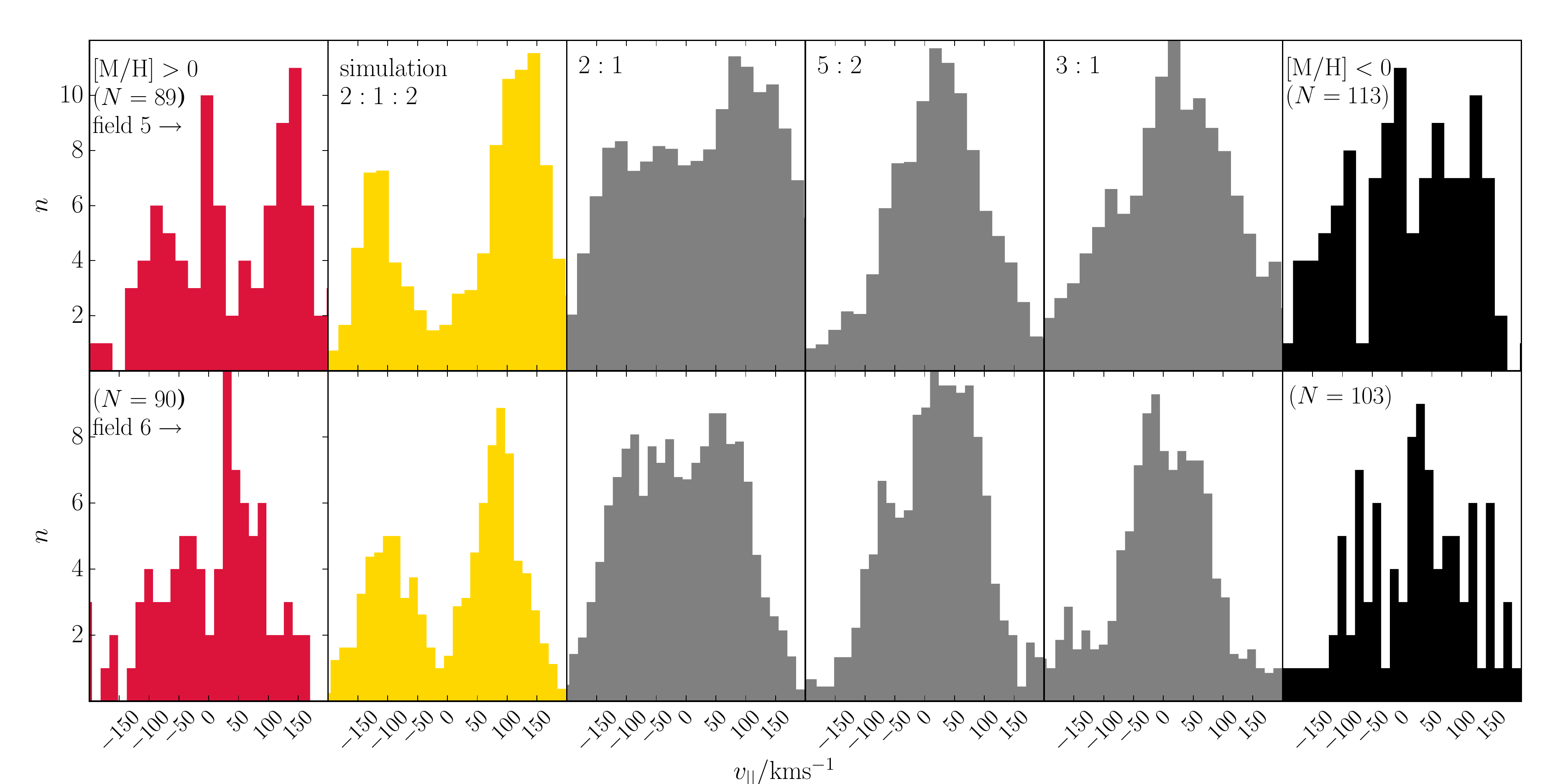}
\caption{\change{Line--of--sight velocity distributions of stars in
    field 5 (Baade's window) and field 6. The red (black) filled
    histograms represent the metal--rich (metal--poor) stars. The
    velocity histograms of some of the dominant resonant orbits from
    the simulation of \citet{Sh10} extracted by the method of
    \citet{Mo15a} are shown in the remaining panels, with the `banana'
    orbits shown in filled yellow. The same bins are used in each row
    of histograms. There is a striking resemblence between the
    velocity peaks in the metal--rich stars and those produced by the
    2:1:2 resonance. The metal--poor stars do not appear to possess
    the same structure.}}
\label{fig:baade}
\end{figure*}

\begin{figure}
\includegraphics[width=.9\columnwidth]{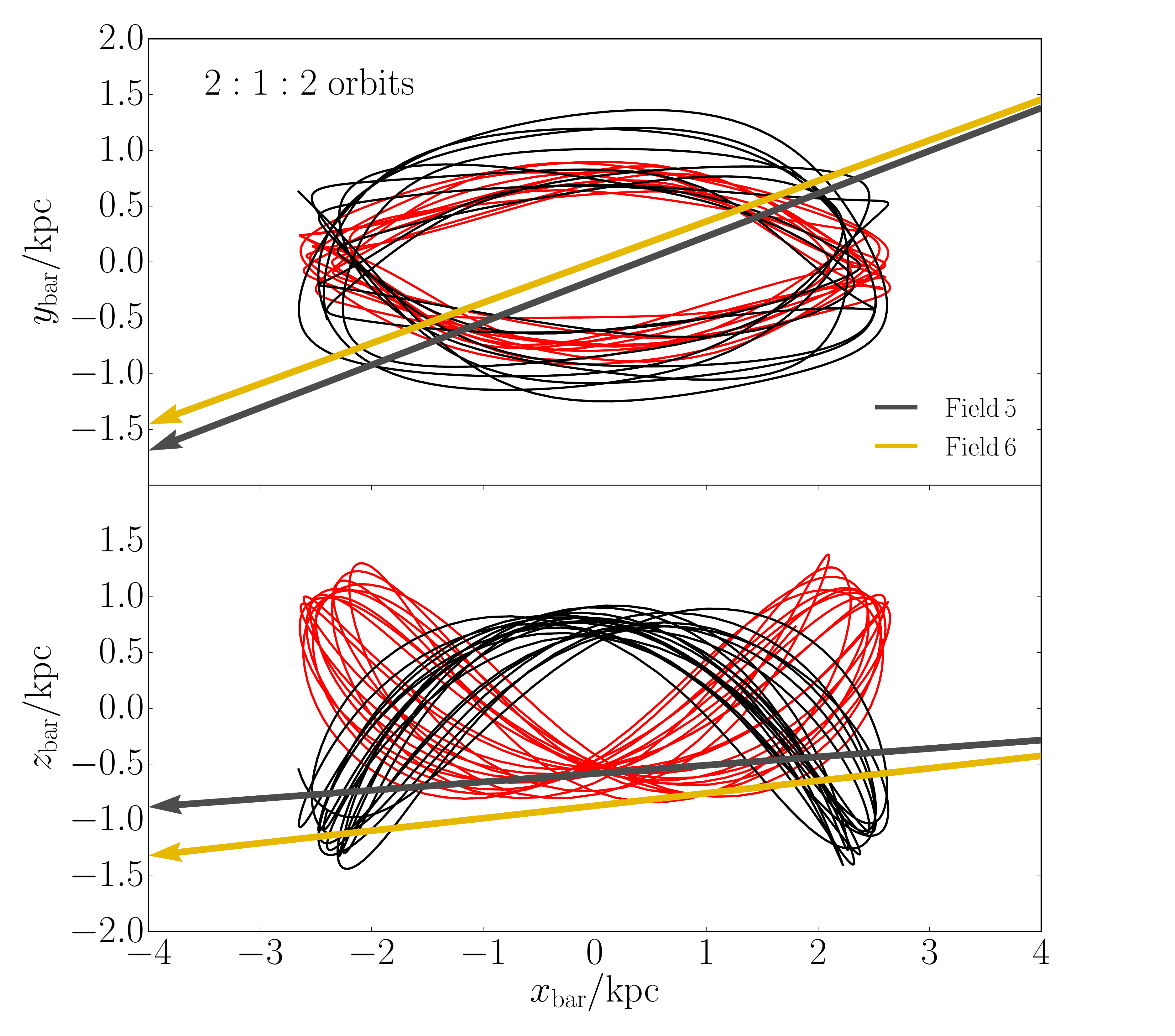}
\caption{Two orbits closely librating around the 2:1:2 resonance in
  the frame of the bar extracted from the simulation of
  \citep{Sh10}. The `banana' shape is clear in the $x-z$ projection. Overplotted 
  are two arrows representing approximate line--of--sight directions for 
  fields 5 (Baade's window, grey) and 6 (gold).}
\label{fig:sampleorbits}
\end{figure}

\begin{figure}
\includegraphics[width=.9\columnwidth]{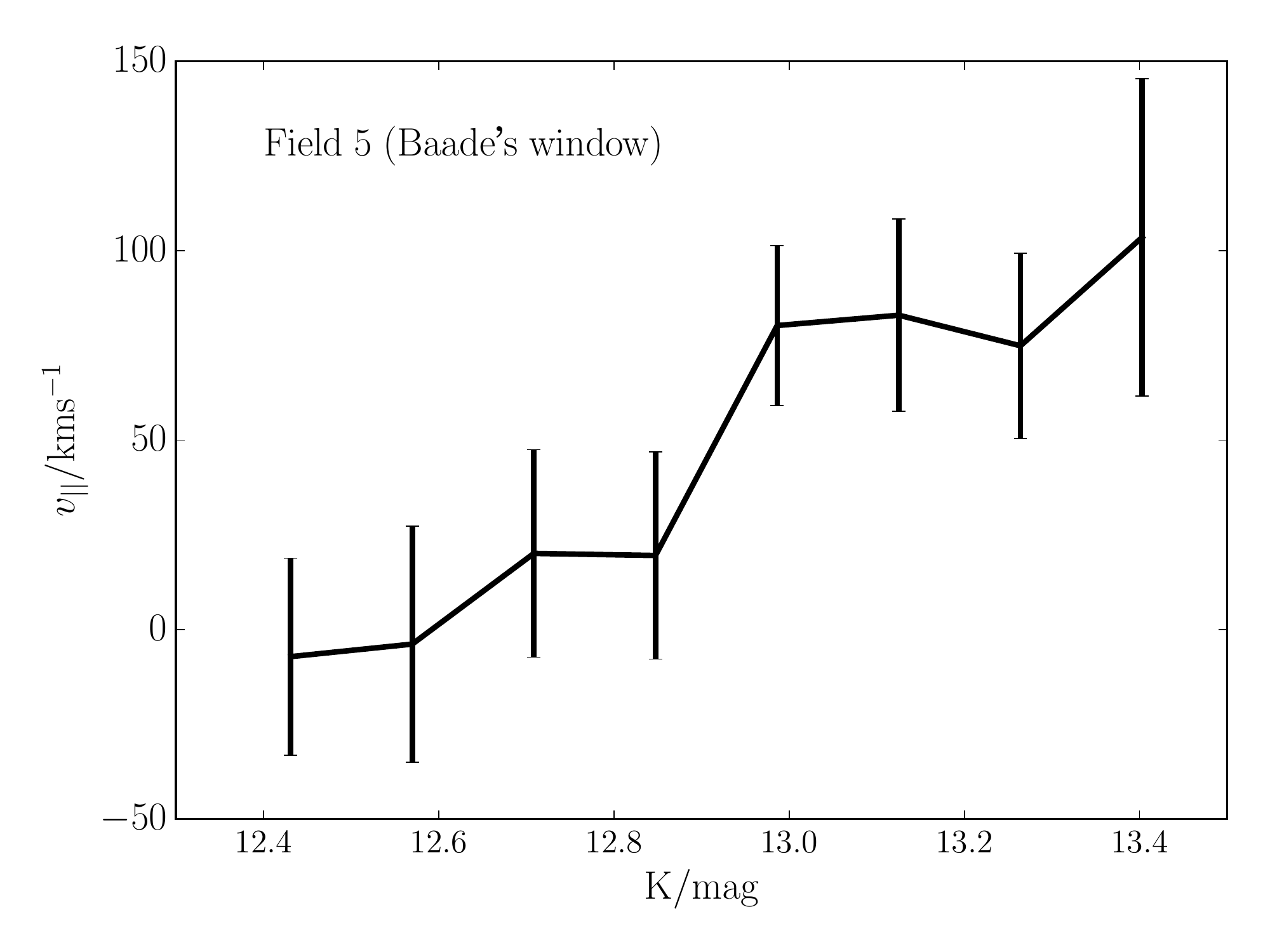}
\caption{The mean line--of--sight velocity of stars in field 5
  (Baade's window) as a function of K--band magnitude. One can see
  evidence for a correlation between K, a rough proxy for distance,
  and the mean line--of--sight velocity. The stars that are on the far
  side of the bulge have larger line--of--sight
  velocities.\change{This is qualitatively compatible with the
    simulation, which predicts a positive correlation between $v_{||}$
    and line--of--sight distance.} Though the signal is noticeable
  here, we do not see an equivalent signal in field 6.}
\label{fig:kmagv}
\end{figure}

Based on the clearly bimodal structure of our sample in the
$\mathrm{[\alpha/M]}$--$\feh$ plane, shown in
Fig. \ref{fig:abunplane}, we split the sample by metallicity. We dub
stars with $\feh>0$ as `metal--rich' and stars with $\feh<0$ as
`metal--poor'. This gives a sample of 1222 metal--poor stars and 716
metal--rich stars. The metal--poor sample is also $\alpha$--rich,
whereas the metal--rich sample is more disc--like ($\alpha$--poor).
We then separately analyse the kinematics of each of these samples
across our twelve fields.  We correct for solar motion using a local
circular speed of $240\kms$ and solar reflex motion of
$(11.1,12.24,7.25)\kms$ \citep{Sc09}.

The Gaia--ESO sample exhibits features that have been observed in
other studies (for a summary, see
\citealt{Ba16}). Fig. \ref{fig:rotation} shows the strong correlation
between mean line--of--sight velocity and Galactic longitude,
indicative of rotation. Here, we find no discernible difference
between the rotational signature in the metal--rich stars and the
metal--poor stars, both of which show a velocity gradient of $\sim
8\kms\deg^{-1}$. The dispersion and shape of the line--of--sight
velocity distributions of each population, however, show marked
differences. Fig. \ref{fig:disps_vs_lat} depicts the line--of--sight 
velocity dispersion of the stars as a function of latitude. 
The metal--rich stars exhibit a steep decline in velocity
dispersion with latitude. A simple linear fit gives a gradient of
$7.4\pm 0.7 \kms\deg^{-1}$. The metal--poor stars, on the other hand,
present a much flatter dispersion profile, with slope $2.0\pm 1.0
\kms\deg^{-1}$. This property has been seen in other surveys
\citep[see Fig. 4 of][]{Ba16}, and has been shown to be consistent
with specific numerical simulations \citep[e.g.,][]{Ba10}.  The
evident separation in both chemistry and kinematics is suggestive that
the two sequences have different formation histories. We \change{might} also
expect interesting or unusual features in the line--of--sight velocity
distributions. For example, \citet{Ni12} first detected cold,
high--velocity peaks in the APOGEE bulge fields near the Galactic
plane, and \citet{Mo15b} subsequently proposed that these structures
were a consequence of stars on resonant orbits in the bar (see also
\citet{De15} for a related idea).

Further examples of this type of phenomenon, now only in the velocity
distributions of the \ges metal rich sample, are shown in the two
leftmost panels of Fig. \ref{fig:baade}. Fields 5 (Baade's window) and
6 are of particular interest because they lie along the minor axis of
the bar, where the X--shape is observed \citep[e.g.,][]{Go15}. They
are also close enough to the Galactic plane that the density of the
boxy--peanut bulge is still relatively high, so the proportion of
metal rich stars is favourable compared to other minor axis fields at
higher latitudes. The velocity distributions for these fields are
multimodal, with narrow peaks at non--zero line--of--sight
velocities. At field 5 (Baade's Window), there are three narrow peaks;
at field 6, there are two broader peaks separated by a small valley.
Such features are present in other fields, and are particularly
prominent at $\ell\sim0^\circ$, along the short--axis of the bar.

\section{Resonant orbits in the bulge}

Orbits in the Galaxy are usually characterised in terms of three
frequencies, $\kappa$, $\Omega$ and $\nu$, describing oscillations in
the radial, azimuthal and vertical directions respectively. The bar
rotates at a pattern speed $\Omega_b$, and this perturbation creates
resonant orbits that satisfy
\begin{equation}
m\kappa = n(\Omega - \Omega_b),
\end{equation} 
where $m$ and $n$ are integers. The quantity $\Omega - \Omega_b$ is
the azimuthal frequency of the resonance in the frame that corotates
with the bar and, since $\Omega - \Omega_b$ is commensurate with
$\kappa$, a resonant orbit will be closed in the corotating frame
(see, e.g. \citealt{Mo15a}).  Simulations of barred spirals as well as
studies of analytic gravitational potentials, confirm that the
morphology of the bar is largely supported by stars librating around
these resonances \citep[e.g.,][]{Mo15b}. The 2:1 resonance (two
radial oscillations per revolution) sires one of the most important
families supporting the backbone of any bar.  Resonances also exist
in which the vertical frequency becomes commensurate with $\kappa$ and
$\Omega - \Omega_b$. `Banana' or $2:1:2$ orbits are examples of this
kind of resonance in which there are also two vertical oscillations per
revolution, as shown in Fig.~\ref{fig:sampleorbits}. 

Here, we will use the sample of resonant orbits extracted by the
method of \citet{Mo15a} from a simulation of \citet{Sh10}. The bar is
formed from $m=2$ instablities in a massive disk which subsequently
become unstable to buckling~\citep[e.g.,][]{Ra91}. The endpoint of the
simulation is a boxy bulge with a pattern speed of $\sim 40$
kms$^{-1}$ kpc$^{-1}$~\citep{Sh16}. When viewed at an angle of
$15^\circ$, the model fits the mean motion and velocity dispersion
data on the minor axis ($b =0^\circ$), as well as a sequence of
longitudinal fields for $b = -4^\circ$ and $b = -6^\circ$ (see Fig.~4
of \citet{Sh10} for more details).  We show as green triangles in
Fig.~\ref{fig:disps_vs_lat} the velocity dispersion of the simulation
particles in the \ges fields as a function of latitude. The simulation
matches the behaviour of the velocity dispersion of the metal-rich
stars.  Although we cannot reasonably expect this simulation to be a
highly accurate match to the inner Milky Way, its structural
properties and gross kinematics are broadly correct. 

There is substantial degeneracy in the way in which different orbits
can be superposed to make a triaxial bar model. The simulation is a
useful tool for understanding which types of orbit are likely
populated, but we do not expect the relative proportion of these
orbits to be correct for the Milky Way bar. \change{If the data set
  were much larger, we could perform Schwarzschild modelling: the
  relative weights of the orbital families extracted from the
  simulation would be fitting parameters. However, since the data are
  sparse, it is unrealistic to expect to constrain
  quantitatively the actual proportions of the different families. Our
  study is less ambitious: we seek to find signs in the data that are
  suggestive of specific orbital families being highly populated.}
Using the method of \citet{Mo15a}, we extract the periodic orbits
lying on the 2:1, 3:1 and 5:2 in-plane resonances from the simulation
by requiring $ (\Omega -\Omega_{\rm p})/\kappa$ to equal $m/n$ within
some tolerance $\epsilon = 0.1$. A heliocentric distance cut of $6.8 <
D/\mathrm{kpc} < 14$ has been applied to the simulation data. The
distance selection function of \ges is much more complicated in
reality, but this is a simple and plausible cut to use when comparing
the simulations to the data.  Finally, we also extract the banana
orbits from the 2:1 sample with the additional cut $0.9 < \nu/\kappa <
1.1$. The velocity histograms provided by these orbital families in
\ges fields 5 and 6 are shown in 
Fig.~\ref{fig:baade}. Field 5 (Baade's window) is relatively well
sampled because of the low reddening. The high velocity peaks in the
data resemble banana orbits moving towards us (predominantly on the
near-side, $v_{||} \sim -90\kms$) and away from us (far-side, $v_{||}
\sim 140\kms$). There is also a narrow peak in the data at
$v_\parallel\approx 0$, which is likely a combination of disk
contamination and other resonances (e.g. 5:2 and 3:1). \change{In
  Field 6, there are possible peaks at $\sim \pm 40\kms$. Finally, we
  note a possibly related result -- a very narrow velocity minimum on
  the Bulge minor axis was also reported in red clump stars in
  Appendix A of \cite{Ne12}}.

If the multi-modal structure of the velocity histogram of metal-rich
stars in Baade's Window is due to banana orbits, then we would expect
to see their signature in other well-sampled minor axis fields. The
second row of Fig.~\ref{fig:baade} shows this is possibly the case. Field
6 is centered on ($\ell = 0.16^\circ, b = -6.03^\circ$). The
bimodality of the velocity histogram data is also matched in the
velocity distribution of the banana orbits. There is, however, a
discrepancy in the location of the velocities of the two peaks. However, this 
is controlled by the curvature of the banana orbits and so may
be adjusted by changing the potential of the bar. In the other minor axis 
\ges fields, the bar density has dropped significantly, and with it the 
number of metal--rich stars in each field. This leads to noisier histograms.

Since we are dealing with relatively few stars, assessment of the
significance of the peaks in the data is necessary. We fitted Gaussian
mixture models to the velocity distributions in fields 5 and 6 using
the {\sc Scikit--learn} module for {\sc Python} \citep{Pe12}. We then
approximated the evidence for a given number of Gaussian components
using the Bayesian information criterion (BIC). The BIC offsets
changes in the maximum likelihood for a given number of Gaussians by a
penalty factor for introducing more parameters. For the metal-poor
stars, only one Gaussian component is favoured in fields 5 and 6,
which is consistent with the hypothesis that they belong to a
dynamically simpler spheroid. For the metal--rich stars, 
two components were favoured in field 5 (Baade's window), so that the
central peak and leftmost peak in the data are combined into a single
Gaussian, and the rightmost peak is the second Gaussian. In field 6,
the BIC informs us that only one Gaussian component is necessary
before over--fitting the metal--rich velocity data.  According to this
particular method, then, the evidence for bimodality in field 6 is
actually weaker than our eyes might have us believe. Nonetheless, we
believe that taken together the velocity distributions of these fields
are interesting. Furthermore, we would expect some level of disc
contamination in both of these fields at around $v_{||}\sim 0 \kms$,
the removal of which could increase the significance of the
bimodality. Assuming the Besan\c{c}on model \citep{Ro03}, the expected level 
of foreground disc contamination in the two fields is $\sim 5$ per
cent. In field 5 (Baade's Window), we find that 11 per cent of stars
have $-10 < v_{||} / \kms < 10$. Given our estimate of the
contamination, a significant fraction of the stars contributing to the
central peak in the velocity histogram of metal--rich stars are likely
disc stars. 

Another intriguing feature in the data is that, in Baade's window, the
line--of--sight velocity and K--band magnitude are correlated (see
Fig. \ref{fig:kmagv}). This is predicted by the simulation, so that
stars in the high velocity peak are found on the far side of the bar
(larger K magnitude). The stars with $v_{||}\sim0\,\kms$ are generally
found at brighter magnitudes, consistent with the picture that they
are contaminants from the disc. Interestingly, we do not see the same
correlation in field 6.

\section{Conclusions}

\change{We compared the velocity histograms of the metal--rich
  ($\feh>0$) stars in the \ges fields on the minor axis of the
  Galactic bulge with those produced by the most populated resonant
  orbits in an N--body simulation of a buckled bar. The narrow peaks
  that are seen in the data resemble those produced by the `banana' or
  2:1:2 resonant orbits in the simulation. Though these results are
  intriguing, more statistics are required to confirm the hypothesis
  that 2:1:2 orbits are heavily populated in the Galactic bar.}

This work supplements the earlier kinematic study of
\citet{Va13}. Targetting the bright and faint red clumps at ($\ell =
0.0^\circ, b = -6.0^\circ$), very close to our field 6, they found
differences in the radial velocity and proper motion distibutions
consistent with stars in the bright clump moving towards us and in the
faint clump moving away. However, they argued that the metal-poor
stars are preferentially on elongated orbits and the metal-rich ones
on more axisymmetric orbits, contrary both to the results in this
paper and theoretical expectation~\citep{Ba10}.  \change{Their
  argument} is open to question as it is not based on orbit
integrations, but on visual inspection of radial velocity histograms
split according to metallicity.  Nonetheless, proper motions may
indeed offer hope for confirming the structure, as the banana orbits
have a characteristic morphology and hence a strong correlation
between $\mu_\ell$, $\mu_b$ and $v_{||}$. \change{The OGLE survey 
has produced such measurements, and {\it Gaia} may also provide proper 
motions for some stars in the bulge.}

The weighting of the different resonances tells us about the
conditions under which the boxy bulge formed via the buckling
instability. Metallicity gradients in the original barred disk will
cause stars with different radii and hence metallicities to be mapped
onto different vertical resonances. This in principle offers us an
opportunity to map out the history of buckling in the inner galaxy, as
well as trace back the state of the pristine disk in the epoch before
buckling.

\acknowledgments

Based on data products from observations made with ESO Telescopes at
the La Silla Paranal Observatory under programme ID 188.B-3002. These
data products have been processed by the Cambridge Astronomy Survey
Unit (CASU) at the Institute of Astronomy, University of Cambridge,
and by the FLAMES/UVES reduction team at INAF/Osservatorio Astrofisico
di Arcetri. These data have been obtained from the Gaia-ESO Survey
Data Archive, prepared and hosted by the Wide Field Astronomy Unit,
Institute for Astronomy, University of Edinburgh, which is funded by
the UK Science and Technology Facilities Council.  This work was
partly supported by the European Union FP7 programme through ERC grant
number 320360 and by the Leverhulme Trust through grant
RPG-2012-541. We acknowledge the support from INAF and Ministero dell'
Istruzione, dell' Universit\`a' e della Ricerca (MIUR) in the form of
the grant "Premiale VLT 2012". The results presented here benefit from
discussions held during the Gaia-ESO workshops and conferences
supported by the ESF (European Science Foundation) through the GREAT
Research Network Programme.

\bibliography{letter}

\begin{thebibliography}{}

\bibitem[\protect\citeauthoryear{{Babusiaux}}{{Babusiaux}}{2016}]{Ba16}
{Babusiaux} C.,  2016, ArXiv e-prints

\bibitem[\protect\citeauthoryear{{Babusiaux}, {G{\'o}mez}, {Hill}, {Royer},
  {Zoccali}, {Arenou}, {Fux}, {Lecureur}, {Schultheis}, {Barbuy}, {Minniti} \&
  {Ortolani}}{{Babusiaux} et~al.}{2010}]{Ba10}
{Babusiaux} C.,  {G{\'o}mez} A.,  {Hill} V.,  {Royer} F.,  {Zoccali} M.,
  {Arenou} F.,  {Fux} R.,  {Lecureur} A.,  {Schultheis} M.,  {Barbuy} B.,
  {Minniti} D.,    {Ortolani} S.,  2010, \aap, 519, A77

\bibitem[\protect\citeauthoryear{{Debattista}, {Ness}, {Earp} \&
  {Cole}}{{Debattista} et~al.}{2015}]{De15}
{Debattista} V.~P.,  {Ness} M.,  {Earp} S.~W.~F.,    {Cole} D.~R.,  2015,
  \apjl, 812, L16

\bibitem[\protect\citeauthoryear{{Gilmore}, {Randich}, {Asplund}, {Binney},
  {Bonifacio}, {Drew}, {Feltzing}, {Ferguson}, {Jeffries}, {Micela} \& et
  al.}{{Gilmore} et~al.}{2012}]{Gi12}
{Gilmore} G.,  {Randich} S.,  {Asplund} M.,  {Binney} J.,  {Bonifacio} P.,
  {Drew} J.,  {Feltzing} S.,  {Ferguson} A.,  {Jeffries} R.,  {Micela} G.,
  et al. 2012, The Messenger, 147, 25

\bibitem[\protect\citeauthoryear{{Gonzalez}, {Zoccali}, {Debattista},
  {Alonso-Garc{\'{\i}}a}, {Valenti} \& {Minniti}}{{Gonzalez}
  et~al.}{2015}]{Go15}
{Gonzalez} O.~A.,  {Zoccali} M.,  {Debattista} V.~P.,  {Alonso-Garc{\'{\i}}a}
  J.,  {Valenti} E.,    {Minniti} D.,  2015, \aap, 583, L5

\bibitem[\protect\citeauthoryear{{McWilliam} \& {Zoccali}}{{McWilliam} \&
  {Zoccali}}{2010}]{McW10}
{McWilliam} A.,  {Zoccali} M.,  2010, \apj, 724, 1491

\bibitem[\protect\citeauthoryear{{Miralda-Escude} \&
  {Schwarzschild}}{{Miralda-Escude} \& {Schwarzschild}}{1989}]{Mi89}
{Miralda-Escude} J.,  {Schwarzschild} M.,  1989, \apj, 339, 752

\bibitem[\protect\citeauthoryear{{Molloy}, {Smith}, {Evans} \& {Shen}}{{Molloy}
  et~al.}{2015}]{Mo15b}
{Molloy} M.,  {Smith} M.~C.,  {Evans} N.~W.,    {Shen} J.,  2015, \apj, 812,
  146

\bibitem[\protect\citeauthoryear{{Molloy}, {Smith}, {Shen} \& {Evans}}{{Molloy}
  et~al.}{2015}]{Mo15a}
{Molloy} M.,  {Smith} M.~C.,  {Shen} J.,    {Evans} N.~W.,  2015, \apj, 804, 80

\bibitem[\protect\citeauthoryear{{Nataf}, {Udalski}, {Gould}, {Fouqu{\'e}} \&
  {Stanek}}{{Nataf} et~al.}{2010}]{Na10}
{Nataf} D.~M.,  {Udalski} A.,  {Gould} A.,  {Fouqu{\'e}} P.,    {Stanek} K.~Z.,
   2010, \apjl, 721, L28

\bibitem[\protect\citeauthoryear{{Ness}, {Freeman}, {Athanassoula},
  {Wylie-de-Boer}, {Bland-Hawthorn}, {Asplund}, {Lewis}, {Yong}, {Lane} \&
  {Kiss}}{{Ness} et~al.}{2013}]{Ne13}
{Ness} M.,  {Freeman} K.,  {Athanassoula} E.,  {Wylie-de-Boer} E.,
  {Bland-Hawthorn} J.,  {Asplund} M.,  {Lewis} G.~F.,  {Yong} D.,  {Lane}
  R.~R.,    {Kiss} L.~L.,  2013, \mnras, 430, 836

\bibitem[\protect\citeauthoryear{{Ness}, {Freeman}, {Athanassoula},
  {Wylie-De-Boer}, {Bland-Hawthorn}, {Lewis}, {Yong}, {Asplund}, {Lane}, {Kiss}
  \& {Ibata}}{{Ness} et~al.}{2012}]{Ne12}
{Ness} M.,  {Freeman} K.,  {Athanassoula} E.,  {Wylie-De-Boer} E.,
  {Bland-Hawthorn} J.,  {Lewis} G.~F.,  {Yong} D.,  {Asplund} M.,  {Lane}
  R.~R.,  {Kiss} L.~L.,    {Ibata} R.,  2012, \apj, 756, 22

\bibitem[\protect\citeauthoryear{{Ness}, {Zasowski}, {Johnson}, {Athanassoula},
  {Majewski}, {Garcia Perez}, {Bird}, {Nidever}, {Schneider}, {Sobeck},
  {Frinchaboy}, {Pan}, {Bizyaev}, {Oravetz} \& {Simmons}}{{Ness}
  et~al.}{2015}]{Ne15}
{Ness} M.,  {Zasowski} G.,  {Johnson} J.~A.,  {Athanassoula} E.,  {Majewski}
  S.~R.,  {Garcia Perez} A.~E.,  {Bird} J.,  {Nidever} D.,  {Schneider} D.~P.,
  {Sobeck} J.,  {Frinchaboy} P.,  {Pan} K.,  {Bizyaev} D.,  {Oravetz} D.,
  {Simmons} A.,  2015, ArXiv e-prints

\bibitem[\protect\citeauthoryear{{Nidever} et~al.,}{{Nidever}
  et~al.}{2012}]{Ni12}
{Nidever} D.~L.,  et~al., 2012, \apjl, 755, L25

\bibitem[\protect\citeauthoryear{{Pedregosa} et~al.,}{{Pedregosa}
  et~al.}{2012}]{Pe12}
{Pedregosa} F.,  et~al., 2012, ArXiv e-prints

\bibitem[\protect\citeauthoryear{{Pfenniger} \& {Friedli}}{{Pfenniger} \&
  {Friedli}}{1991}]{Pf91}
{Pfenniger} D.,  {Friedli} D.,  1991, \aap, 252, 75

\bibitem[\protect\citeauthoryear{{Portail}, {Wegg} \& {Gerhard}}{{Portail}
  et~al.}{2015}]{Po15}
{Portail} M.,  {Wegg} C.,    {Gerhard} O.,  2015, \mnras, 450, L66

\bibitem[\protect\citeauthoryear{{Raha}, {Sellwood}, {James} \& {Kahn}}{{Raha}
  et~al.}{1991}]{Ra91}
{Raha} N.,  {Sellwood} J.~A.,  {James} R.~A.,    {Kahn} F.~D.,  1991, \nat,
  352, 411

\bibitem[\protect\citeauthoryear{{Robin}, {Reyl{\'e}}, {Derri{\`e}re} \&
  {Picaud}}{{Robin} et~al.}{2003}]{Ro03}
{Robin} A.~C.,  {Reyl{\'e}} C.,  {Derri{\`e}re} S.,    {Picaud} S.,  2003,
  \aap, 409, 523

\bibitem[\protect\citeauthoryear{{Rojas-Arriagada} et~al.,}{{Rojas-Arriagada}
  et~al.}{2014}]{Ro14}
{Rojas-Arriagada} A.,  et~al., 2014, \aap, 569, A103

\bibitem[\protect\citeauthoryear{{Sch{\"o}nrich}, {Binney} \&
  {Dehnen}}{{Sch{\"o}nrich} et~al.}{2010}]{Sc09}
{Sch{\"o}nrich} R.,  {Binney} J.,    {Dehnen} W.,  2010, \mnras, 403, 1829

\bibitem[\protect\citeauthoryear{{Shen} \& {Li}}{{Shen} \& {Li}}{2016}]{Sh16}
{Shen} J.,  {Li} Z.-Y.,  2016, Galactic Bulges, 418, 233

\bibitem[\protect\citeauthoryear{{Shen}, {Rich}, {Kormendy}, {Howard}, {De
  Propris} \& {Kunder}}{{Shen} et~al.}{2010}]{Sh10}
{Shen} J.,  {Rich} R.~M.,  {Kormendy} J.,  {Howard} C.~D.,  {De Propris} R.,
  {Kunder} A.,  2010, \apjl, 720, L72

\bibitem[\protect\citeauthoryear{{V{\'a}squez}, {Zoccali}, {Hill}, {Renzini},
  {Gonz{\'a}lez}, {Gardner}, {Debattista}, {Robin}, {Rejkuba}, {Baffico},
  {Monelli}, {Motta} \& {Minniti}}{{V{\'a}squez} et~al.}{2013}]{Va13}
{V{\'a}squez} S.,  {Zoccali} M.,  {Hill} V.,  {Renzini} A.,  {Gonz{\'a}lez}
  O.~A.,  {Gardner} E.,  {Debattista} V.~P.,  {Robin} A.~C.,  {Rejkuba} M.,
  {Baffico} M.,  {Monelli} M.,  {Motta} V.,    {Minniti} D.,  2013, \aap, 555,
  A91

\end{thebibliography}
\bibliographystyle{mn2e}

\end{document}